\documentclass[12pt]{article}

\makeatletter
\renewcommand\section{\@startsection {section}{1}{\z@}%
                                   {-3.5ex \@plus -1ex \@minus -.2ex}
                                   {2.3ex \@plus.2ex}%
                                   {\normalfont\large\bfseries}}
\renewcommand\subsection{\@startsection{subsection}{2}{\z@}%
                                     {-3.25ex\@plus -1ex \@minus -.2ex}%
                                     {1.5ex \@plus .2ex}%
                                     {\normalfont\bfseries}}
\makeatother

\def\IZ{\relax\ifmmode\mathchoice
{\hbox{\cmss Z\kern-.4em Z}}{\hbox{\cmss Z\kern-.4em Z}}
{\lower.9pt\hbox{\cmsss Z\kern-.4em Z}} {\lower1.2pt\hbox{\cmsss
Z\kern-.4em Z}}\else{\cmss Z\kern-.4em Z}\fi}
\def\IR{\relax{\rm I\kern-.18em R}}

\def\one{{\hbox{ 1\kern-.8mm l}}}

\newlength{\bredde}
\def\slash#1{\settowidth{\bredde}{$#1$}\ifmmode\,\raisebox{.15ex}{/}
\hspace*{-\bredde} #1\else$\,\raisebox{.15ex}{/}\hspace*{-\bredde}
#1$\fi}

\newsavebox{\zzzbar}
\sbox{\zzzbar}
  {\setlength{\unitlength}{0.9em}
  \begin{picture}(0.6,0.7)
  \thinlines
  \put(0,0){\line(1,0){0.6}}
  \put(0,0.75){\line(1,0){0.575}}
  \multiput(0,0)(0.0125,0.025){30}{\rule{0.3pt}{0.3pt}}
  \multiput(0.2,0)(0.0125,0.025){30}{\rule{0.3pt}{0.3pt}}
  \put(0,0.75){\line(0,-1){0.15}}
  \put(0.015,0.75){\line(0,-1){0.1}}
  \put(0.03,0.75){\line(0,-1){0.075}}
  \put(0.045,0.75){\line(0,-1){0.05}}
  \put(0.05,0.75){\line(0,-1){0.025}}
  \put(0.6,0){\line(0,1){0.15}}
  \put(0.585,0){\line(0,1){0.1}}
  \put(0.57,0){\line(0,1){0.075}}
  \put(0.555,0){\line(0,1){0.05}}
  \put(0.55,0){\line(0,1){0.025}}
  \end{picture}}

\newcommand{\ena}{\end{eqnarray}}
\newcommand{\beqa}{\begin{eqnarray}}
\newcommand{\eeqa}{\end{eqnarray}}
\newcommand{\bea}{\begin{eqnarray}}
\newcommand{\eea}{\end{eqnarray}}

\newcommand{\eq}[1]{(\ref{#1})}

\newcommand{\be}{\begin{equation}}
\newcommand{\ee}{\end{equation}}

\usepackage{graphicx}

\newcommand{\beq}{\begin{equation}}
\newcommand{\eeq}{\end{equation}}
\newcommand{\ber}{\begin{array}}
\newcommand{\eer}{\end{array}}

\newcommand{\del}{\partial}

\newcommand{\dsty}{\displaystyle}

\newcommand{\te}{\theta}

\newcommand{\de}{\delta}

\newcommand{\eps}{\varepsilon}

\begin{document}
\begin{titlepage}
\begin{flushright}
hep-th/0608123, ITFA-2006-27
\end{flushright}
\vfill
\begin{center}
{\LARGE\bf Local recoil of extended solitons:\vspace{4mm}\\a string theory example}    \\
\vskip 10mm
{\large Ben Craps,$^{a,b}$ Oleg Evnin$^{c,d}$ and Shin Nakamura$^{e,f}$}
\vskip 7mm
{\em $^{a}$ Theoretische Natuurkunde, Vrije Universiteit Brussel and\\
The International Solvay Institutes\\ Pleinlaan 2, B-1050 Brussels, Belgium\footnote{present address}}
\vskip 3mm 
{\em $^b$ Instituut voor Theoretische Fysica, Universiteit van Amsterdam, Valckenierstraat 65, 1018 XE Amsterdam, The Netherlands}
\vskip 3mm
{\em $^c$ California Institute of Technology 452-48, Pasadena, CA 91125, USA}  
\vskip 3mm
{\em $^d$ Department of Physics, Rostov State University, pr. Zorge 5,\\344014 Rostov-na-Donu, Russia}  
\vskip 3mm
{\em $^e$ Physics Department, Hanyang University, Seoul, 133-791, Korea}
\vskip 3mm
{\em $^f$ Center for Quantum Spacetime (CQUeST), Sogang University, Seoul, 121-742, Korea}
\vskip 3mm
{\small\noindent  {\tt Ben.Craps@vub.ac.be, eoe@caltech.edu, nakamura@hanyang.ac.kr}}
\end{center}
\vfill

\begin{center}
{\bf ABSTRACT}
\end{center}
It is well-known that localized topological defects (solitons) experience recoil when they suffer an impact by incident particles. Higher-dimensional topological defects develop distinctive wave patterns propagating along their worldvolume under similar circumstances. For 1-dimensional topological defects (vortex lines), these wave patterns fail to decay in the asymptotic future: the propagating wave eventually displaces the vortex line a finite distance away from its original position (the distance is proportional to the transferred momentum). The quantum version of this phenomenon, which we call ``local recoil'', can be seen as a simple geometric manifestation of the absence of spontaneous symmetry breaking in 1+1 dimensions. Analogously to soliton recoil,
local recoil of vortex lines is associated with infrared divergences in perturbative expansions. In perturbative string theory, such divergences appear in amplitudes for closed strings scattering off a static D1-brane. Through a Dirac-Born-Infeld analysis, it is possible to resum these divergences in a way that yields finite, momentum-conserving amplitudes. 
\vfill

\end{titlepage}
\section{Introduction}

The phenomenon of soliton recoil has been familiar for a few decades by now \cite{rajaraman}. Solitons in quantum field theory are quantum descendants of topologically non-trivial solutions of the classical field equations. In attempting to examine the scattering of fundamental field quanta in the background of (localized) solitons, one discovers infrared divergences in perturbation theory. These divergences signify a need for a background modification. This is not surprising, since solitons have a finite mass and necessarily start moving as a result of the impact by the incident particles. The original classical solution does not properly accommodate this aspect of scattering dynamics. A string theory version of this phenomenon is D0-brane recoil under the impact of incident closed strings. Indeed, if one starts with a background of a D0-brane at rest, modular integrals over worldsheets with holes exhibit divergences, thus signaling the need for a background modification. This background modification precisely corresponds to D0-brane recoil.

Infinitely extended solitons cannot recoil homogeneously due to their infinite mass. Nevertheless, by energy-momentum conservation, during the impact, the incident particles necessarily transfer to the extended soliton a certain amount of momentum in the transverse directions. This momentum influx induces a wave-like perturbation propagating along the worldvolume. For extended solitons with more than one non-compact spatial dimension, each point returns to its original position after the wave has passed. In fact, a smooth localized impact induces approximately spherical waves that decay as they propagate to infinity.

The fact that a localized impact induces only a small transient disturbance on the worldvolume of extended solitons may make one expect that infrared divergences (of the kind associated with the recoil of localized solitons) are absent from amplitudes for scattering off extended solitons. This is indeed the case for extended solitons with more than one non-compact spatial dimension.
However, for one-dimensional extended solitons (such as vortex lines in four space-time dimensions), one discovers infrared divergences, much in the same way
as for localized solitons. One is then faced with a need to give a physical
interpretation to such divergences, and to show how they should be handled in
perturbation theory.

\begin{figure}[t]
\hspace{2cm}\includegraphics[height=8cm]{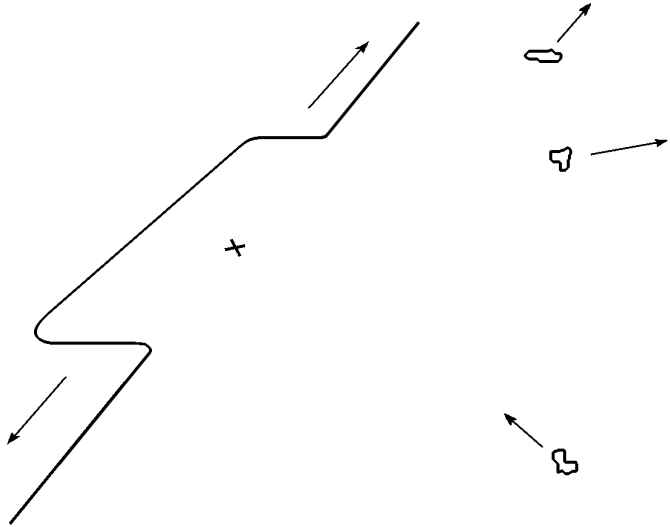}
\caption{A pictorial representation of local recoil. The closed strings scattering off a D1-brane induce two waves propagating away from the point of impact (denoted by {\tt x}).}
\label{localrecoil}
\end{figure}

The disturbance produced by the impact of the incident particles behaves in a very different way for vortex lines and the higher-dimensional solitons. For vortex lines, the effect of a delta-function kick at $t=0$ and $x=0$ is given by the retarded Green function of the (1+1)-dimensional wave operator describing the propagation of small transverse perturbations:
\beq
G_{(1+1)}(t,x)\sim \theta (t-|x|),
\label{green}
\eeq
where $\theta(x)$ is the step function. Quite obviously, the perturbation does not decay for large times. Rather, the system exhibits two kink-like waves propagating towards infinity and shifting the vortex line by a finite distance off its initial position as they pass along (the distance is proportional to the momentum transferred to the vortex line). It is this large-scale, non-transient response of vortex lines to the impact by incident particles that is responsible for the infrared divergences present in perturbative expansions of scattering amplitudes in the background of static vortex lines. A string theory version of this phenomenon is the scattering of closed strings off a D1-brane.%
\footnote{
Transverse fluctuations of the D1-brane worldsheet are described by massless scalar fields, arising from quantizing open strings ending on the D1-brane. In a low energy limit, their dynamics is given by a canonical kinetic term, giving rise to the retarded Green function \eq{green}. Strictly speaking, the low energy approximation would break down for a delta-function impact, as illustrated by the unphysical instantaneous displacement of each point of the string in \eq{green}. However, we have in mind an appropriate superposition representing an impact smoothed out over, for instance, a string length, for which the approximation would be valid. The conclusion that the whole string ends up displaced by a finite distance continues to hold. 
}
A pictorial representation of this process is given in Fig.~\ref{localrecoil}. We term it ``local recoil''.

One may wonder why the local recoil wave does not decay into radiation, as would happen, for example, to oscillation modes of macroscopic strings in a cosmological context. One intuitive explanation invokes the non-oscillating character of the local recoil wave. Moreover, because of the large tension of the D1-brane, inversely proportional to the string coupling constant $g_{st}$, the kinematics implies that the momentum 
transferred to the D1-brane is space-like%
\footnote{
As in the previous footnote, we are assuming the impact to be smeared over a distance of order the string length, so that the relevant portion of the D1-brane behaves as a very heavy particle, which absorbs momentum more easily than energy.
}
and thus cannot be (completely) absorbed by any assortment of outgoing particles.

The purely kinematic nature of the local recoil phenomenon makes it universal. One should expect that a generic vortex line configuration will undergo local recoil whenever particles scatter off it, to a great extent irrespectively of the dynamical content of the theory. In appendix~A, we show that recoil-related infrared divergences are indeed present for vortex lines in a field theory example.

In what follows, we shall examine the local recoil of a (bosonic) D1-brane in some detail. It will be shown how infrared divergences signalling the need for a background modification arise, and how they can be resummed in a way that enforces momentum conservation. 

At this point, we should warn the reader that the above pictorial representation, where a localized D1-brane undergoes local recoil, holds only in the classical approximation. Because of large quantum fluctuations in two dimensions, the quantum theory of a D1-brane does not even have localized ground states. In fact, it will turn out that the state implicitly chosen in perturbative string theory is completely delocalized. Still, a quantum version of the local recoil phenomenon exists.

\section{The annular divergence}

The non-perturbatively large tension of D-branes ($\sim 1/g_{st}$) makes any background modification unnecessary to lowest order in the string coupling, when a D-brane is hit by closed strings (whose momenta are kept finite as $g_{st}$ goes to zero). Infrared divergences (and the corresponding need for a background modification) may arise however if one tries to compute the next-to-leading order corrections to the scattering amplitude. In string theory, these next-to-leading order corrections come from string worldsheets of annular topology.

In field theory, the relevant infrared divergences in loop diagrams come from large distance propagation of the zero modes corresponding to displacing the topological defect. In string theory, such large distance propagation corresponds to the annulus developing a long, thin strip.

Divergences from degenerating Riemann surfaces can be analyzed using Polchinski's plumbing fixture construction \cite{polchinski-fischler-susskind}, which relates the divergences to amplitudes evaluated on a lower genus Riemann surface. In particular, the annulus amplitude with an insertion of vertex operators $V^{(1)},\cdots,V^{(n)}$ (in the interior) can be expressed through disk amplitudes with additional operator insertions at the boundary:
\beq
\left\langle V^{(1)}\cdots V^{(n)}\right>_{\mbox{\small annulus}}=\sum\limits_\alpha \int \frac{dq}q\, q^{h_\alpha-1} \int d\te d\te' \left\langle V_\alpha(\te)V_\alpha(\te')V^{(1)}\cdots V^{(n)}\right>_{D_2},
\label{plumbing}
\eeq
where the summation extends over a compete set of local operators $V_\alpha(\te)$ with conformal weights $h_\alpha$, and $q$ is the gluing parameter that can be related to the annular modulus. ($\te$ parametrizes the boundary of the disk.) The divergence in the integral over $q$ coming from the region $q\approx0$ (i.e., from an annulus developing a thin strip) will be dominated by the terms with the smallest possible $h_\alpha$. 

Let us consider the annular divergence in a somewhat more general setting, when closed strings scatter off a static $Dp$-brane with $d$ non-compact Neumann directions, $p+1-d$ compact Neumann directions and $25-p$ (non-compact) Dirichlet directions. Neglecting the tachyon divergence, which is a pathology peculiar to the case of the bosonic string, we consider the following operators with conformal weights $h=1+\alpha'\kappa^2+4\alpha'\pi^2\sum n_r^2/L_r ^2$:
\be
V^i(\te,\kappa^a,n^r )=\ :\del_n X^i(\te)\exp\left[i\kappa^a X^a(\te)\right] \exp\left[\frac{2\pi i n^r  X^r (\te)}{L_r }\right]:\ .
\ee
These operators correspond to massless open string states (corresponding to fluctuations of the D-brane in the $i$'th Dirichlet direction) carrying momentum $\kappa^a$ in the non-compact Neumann directions and $n^r $ units of Kaluza-Klein momentum in the $r $'th compact Neumann direction. $L_r/2\pi$ are the compactification radii and $\del_n$ denotes the normal derivative. 

It is easy to see that only the operators with $n_r =0$ will contribute to the leading divergence in (\ref{plumbing}). Furthermore, for small values of $q$ (which is the region we are interested in), only small values of $\kappa^a$ will contribute to the integral. With these specifications, we can transform the annular divergence as follows:
\bea
&&\left\langle V^{(1)}\cdots V^{(n)}\right>_{\mbox{\small annulus}}\nonumber\\
&&\sim\int\limits_0^1 dq\int  
d\vec{\kappa}\:
q^{-1+\alpha'\kappa^2}\int d\te d\te' 
\left\langle V^i(\te,\kappa^a,0)V^i(\te',\kappa^a,0)V^{(1)}\cdots V^{(n)}\right>_{D_2}\nonumber\\
&&\sim\int\limits_0^1 dq \int 
d\vec{\kappa}\:
q^{-1+\alpha'\kappa^2}\int d\te d\te' \left\langle V^i(\te,0,0)V^i(\te',0,0)V^{(1)}\cdots V^{(n)}\right>_{D_2}\nonumber\\
&&\sim P^2\left\langle V^{(1)}\cdots V^{(n)}\right>_{D_2}\int\limits_0^1 dq\int 
d\vec{\kappa}\:
q^{-1+\alpha'\kappa^2},
\label{annulus}
\eea
where in the last step we have taken into account the fact that the operator $\int \del_n X^i(\te)d\te$ merely shifts the position of the D-brane; inserting it into any amplitude amounts to multiplication by the total (Dirichlet) momentum $P$ transferred by the closed strings to the D-brane during scattering. We further notice that
\beq
\int\limits_0^1 dq \int 
d\vec{\kappa}\:
q^{-1+\alpha'\kappa^2}\sim\int\limits_0^1 dq \int\limits_0^\infty d\kappa \kappa^{d-1} q^{-1+\alpha'\kappa^2}\sim\int\limits_0^1 \frac{dq}{q\,(\log q)^{d/2}}\ ,
\label{q}
\eeq
where, once again, $d$ is the number of non-compact Neumann directions.%
\footnote{
Note that the vertex operators corresponding to the D-brane gauge field do not contribute to divergence associated to recoil. This is due to the fact that the operator $\int d\te :\del_\te X^i(\te)\exp\left[i\kappa^a X^a(\te)\right]:$ vanishes as $\kappa^a\to 0$ (unless the closed string vertex operators transfer string winding number to the D-brane, which is possible if some of the Neumann directions are compactified; in this paper we only consider divergences due to momentum transfer, though). 
}

Divergences from the region $q\sim0$ in the above integral are only present for the low-dimensional cases. For $d=0$, i.e., a D-instanton, introducing a cut-off $\eps$ on the lower bound of the 
integral (\ref{q}) reveals a $\log\eps$ divergence first described in \cite{combinatorics}. The divergence for $d=1$, i.e., a D0-brane, is $\sqrt{|\log\eps|}$. It is indicative of recoil \cite{tafjord, fischler}. For $d=2$, we observe a $\log|\log\eps|$ divergence which is a manifestation of local recoil, the subject of the present paper. There are no annular divergences for $d>2$, which is consistent with the comments in the introduction. 

\section{The limited role of the Fischler-Susskind mechanism}

For the case of D0-branes, the annulus divergence is indicative of the failure of the worldsheet conformal field theory (CFT) to describe the translational motion of the D0-brane, and it is most properly dealt with by introducing the corresponding collective coordinate explicitly, as per construction of the worldline formalism. This formalism was first introduced in \cite{hirano-kazama}, further developed in \cite{Evnin:2005td}, and will be used for a detailed study of D0-brane recoil in \cite{CEN}. Alternatively, to lowest order in the string coupling, one could deform the worldsheet CFT by an appropriate recoil operator, so that the annular divergence is cancelled and the final state of the D0-brane is changed in such a way that momentum is conserved \cite{CEN} (see \cite{tafjord, fischler} for earlier proposals). This
latter approach is an open string analog of the Fischler-Susskind
mechanism of divergence cancellation in the closed string sector \cite{fischler-susskind}.

Does this strategy translate in a meaningful way to the case of a D1-brane? The most practical answer is no, and the reason is twofold. First, unlike for the case of a D0-brane \cite{CEN}, for a D1-brane one can explicitly specify its vibrational state by including the corresponding incoming and outgoing open strings in the scattering amplitude. Second, unlike for the case of the D0-brane, there is no unique momentum-conserving final state of the D1-brane. Hence, even if some sort of recoil operator is introduced, one would still need to specify the final vibrational state of the D1-brane in terms of open strings. Deforming the CFT with a recoil operator will merely re-shuffle the assignment of the various open string states to the various vibrational modes of the D1-brane.

What is then the appropriate way to deal with the double-log annular divergence present in the background of a D1-brane? One could in principle construct (to lowest order in $g_{st}$) the various assortments of open strings that carry the appropriate amount of Dirichlet momentum and show that the amplitude to scatter into such states is manifestly finite. However, as we shall see in the next section, an even clearer picture of the inner workings of the divergence cancellation emerges if such computations are performed at the level of the Dirac-Born-Infeld (DBI) action (with the momenta of the incident closed string restricted to values much smaller than the string scale). Indeed, the DBI formalism allows to work with the infrared divergences to all orders in the string coupling, and, as a result, it becomes apparent that the infrared divergences in the worldsheet computations arise from attempting to expand in a Taylor series the non-analytic dependences appearing in the scattering amplitudes. Such non-analytic momentum dependences are a direct consequence of momentum conservation (and a direct analog of the momentum conservation $\de$-function appearing for D0-branes \cite{CEN}). The presence of the infrared divergences in the CFT formalism is thereby related to the physical underpinnings of the local recoil phenomenon.

\section{Effective field theory, D1-brane final state and divergence cancellation}

The effective field theory description of D-branes is a very powerful framework to study the infrared divergences associated with recoil, because, as we will see, the non-polynomial form of the DBI action effectively leads to a resummation of infrared divergences at all orders in the string coupling. A drawback of the effective field theory approach is that the momenta of the incident closed strings must be restricted to values much smaller than the string scale. Considering this kinematic region is nevertheless sufficient to corroborate the qualitative description of the local recoil outlined in the introduction.

We start with the familiar DBI action for a D$p$-brane:
\be
S_{DBI}=-\tau\int d^{p+1}\xi\,e^{-\Phi(X(\xi))}\left\{\det\left[\frac{\del X^\mu}{\del\xi^a}\frac{\del X^\nu}{\del\xi^b}\left(G_{\mu\nu}(X(\xi))+B_{\mu\nu}(X(\xi))\right)+2\pi\alpha'F_{ab}\right]\right\}^{1/2}
\ee
and restrict ourselves, for concreteness, to scattering of one dilaton off an initially static, flat, infinitely extended D1-brane in Minkowski space. In this case, the $G$ and $B$ fields can be set to their (Minkowski) background values. Since the energy of the incident closed strings is taken to be small, the D1-brane will be perturbed off its stationary configuration only slightly:
\be
X^0(x,t)=t\qquad X^1(x,t)=x\qquad X^i(x,t)=Y^i(x,t)
\ee
(we are considering a D1-brane stretched in the first spatial direction, and the Dirichlet directions are labelled by $i$). As we remarked in section 2, the D-brane gauge fields do not contribute to the infrared divergences associated to recoil. At the level of effective field theory, this originates from the fact that the gauge fields only interact with (non-winding) closed string fields through terms in the Lagrangian that contain derivatives in the Neumann directions. Such derivatives soften the infrared behavior in loops. We therefore omit the gauge fields from the Lagrangian for the purposes of our analysis. For one dilaton scattering off the D1-brane, the relevant part of the Lagrangian is
\beq
\tau\int dt\,dx \left\{\frac12 \left(\del_t Y\right)^2 - \frac12 \left(\del_x Y\right)^2 - \Phi\left(t,x,Y^i(t,x)\right) + \frac12\left(\Phi(t,x,Y^i(t,x))\right)^2\right\},
\label{DBI_D1}
\eeq
where $\tau$ is the tension of the D1-brane. The interactions of the $Y$-scalars involving derivatives in the Neumann directions have been omitted as they do not contribute to the infrared divergences (in analogy to the gauge fields). 

To analyze the pattern of infrared divergence resummation, let us examine the contribution to the dilaton scattering amplitude from the last term in (\ref{DBI_D1}). In operator language (we work in the interaction picture), it is
\beq
\left\langle f,k_2\right|\frac{\tau}2\int dt dx \left(\Phi(t,x,Y^i(t,x))\right)^2\left|i,k_1\right>,
\label{D1amplitude}
\eeq
where $\left\langle f\right|$ and $\left|i\right>$ describe the initial and final state of the D1-brane, and $\left\langle k_2\right|$ and $\left|k_1\right>$ describe the outgoing dilaton of momentum $k_2$ and incoming dilaton of momentum $k_1$. Using
\be\label{phisquared}
\left\langle k_2\right|\left(\Phi(t,x,y^i)\right)^2\left|k_1\right>\sim\exp\left[i(k_2^0-k_1^0)t\right]\exp\left[-i(k_2^1-k_1^1)x\right] \exp\left[-i(k_2^i-k_1^i)y^i\right]
\ee
(we will not keep track of the overall coefficient of the amplitude), (\ref{D1amplitude}) can be rewritten as
\beq
\left\langle f\right|\int dx dt \exp\left[-i(k_2^i-k_1^i)Y^i(t,x)\right]\exp\left[i(k_2^0-k_1^0)t\right] \exp\left[-i(k_2^1-k_1^1)x\right]\left|i\right>.
\label{D1matrix}
\eeq

Let us choose $|i\rangle$ to be the vacuum state $|0\rangle$ of the stretched D1-brane, and let $|f\rangle$ be 
a coherent state\footnote{
For another use of coherent states in the context of D0-brane 
recoil, see \cite{Shin}.
} 
corresponding to the complex amplitude $v^i(\kappa)$ for the D1-brane oscillation of momentum $\kappa$:
\be
\left|v(\kappa)\right>=
\exp\left[\int d\kappa\left(v^i(\kappa) a^{i\dagger}(\kappa)-\overline{v^i(\kappa)} a^{i}(\kappa)\right)\right]|0\rangle,
\ee
where $a^{i\dagger}$ and $a^i$ are creation-annihilation operators corresponding to the fluctuations of the D1-brane:
\be
Y^i(t,x)=\frac{1}{\sqrt{2\pi\tau}}\int \frac{d\kappa}{\sqrt{2|\kappa|}}\left(a^i(\kappa)e^{i(\kappa x-|\kappa|t)}+a^{i\dagger}(\kappa)e^{-i(\kappa x-|\kappa|t)}\right).
\ee
Since the exponential of the field creates a coherent state, we can rewrite (\ref{D1matrix}) as
\beq
\int dx dt \exp\left[i(k_2^0-k_1^0)t\right] \exp\left[-i(k_2^1-k_1^1)x\right]\left\langle v(\kappa) \Biggl|-\frac{i(k_2^i-k_1^i)}{2\sqrt{\pi\tau|\kappa|}}e^{-i(\kappa x-|\kappa|t)}\right\rangle ,
\label{coherent}
\eeq
where both the bra and the ket denote coherent states. Evaluating the inner product of the two coherent states, (\ref{coherent}) becomes
\bea
&&\int dx dt \exp\left[i(k_2^0-k_1^0)t\right]\exp\left[-i(k_2^1-k_1^1)x\right]\nonumber\\
&&\times\exp\left[-\frac12\int d\kappa\left(|v^i(\kappa)|^2+\frac{(k_2^i-k_1^i)^2}{4\pi\tau|\kappa|}+\frac{2i \overline{v^i(\kappa)}(k_2^i-k_1^i)e^{-i(\kappa x-|\kappa|t)}}{2\sqrt{\pi\tau|\kappa|}}\right)\right].
\label{inprodcoh}
\eea
It is important to note that the integral in the exponential in the second line of \eq{inprodcoh} diverges at small $\kappa$ for generic $v(\kappa)$. This can be understood from the fact that, for generic $v(\kappa)$, the final state of the D1-brane does not satisfy momentum conservation, and the corresponding amplitude should vanish.%
\footnote{\label{footnote:higher}
Note that for higher-dimensional branes, the small $\kappa$ divergence is absent. As we will see in the next section, this is related to the spontaneous breaking of the translational invariance in the Dirichlet directions for higher-dimensional branes.}

To make this more explicit, let us write
\be
v^i(\kappa)=\frac{A^i}{\sqrt{|\kappa}|}+\tilde v^i(\kappa),
\ee
where $A^i$ are constants
and $\tilde v(\kappa)$ is less singular than $1/{\sqrt{|\kappa|}}$ as $\kappa$ goes to 0. The amplitude (\ref{coherent}) then vanishes unless
\beq
A^i=\frac{i(k_1^i-k_2^i)}{2\sqrt{\pi\tau}}.
\label{Ai}
\eeq
The meaning of this condition can be clarified by considering the Dirichlet momentum carried by the D1-brane:
\beq
P^i=\tau\int dx\,\del_t Y^i(x,t)=-i\lim\limits_{\kappa\to 0}\sqrt{\pi\tau|\kappa|}\,(a^i(\kappa)-a^{i\dagger}(\kappa)). 
\label{Dmomentum}
\eeq
If we now compute the expectation value of $P$ for our final state $|v(\kappa)\rangle$, we get
\be
\langle v(\kappa)|P^i|v(\kappa)\rangle=-i\sqrt{\pi\tau}\left(A^i-\bar A^i\right).
\ee
Therefore, (\ref{Ai}) is nothing but the momentum conservation condition:
\be
P+k_2-k_1=0.
\ee
In other words, the scattering amplitude (\ref{D1amplitude}) vanishes unless the momentum conservation condition is satisfied.%
\footnote{
The fact that the amplitude vanishes unless $P+k_2-k_1=0$ strongly suggests
that the state $|v(k)\rangle$ has a definite value of the Dirichlet momentum
$P$. A rigorous proof of this statement would require a careful treatment
of the large wavelength sector of the quantum string Hilbert space, in particular, on account of the $\kappa\to0$ limit in the definition of $P$.
(The formal difficulties with quantizing a massless (1+1)-dimensional scalar
field have been known for a long while.)
}

Can we relate the property that the amplitude vanishes unless the momentum conservation condition is satisfied to the annular divergences in the string diagrams found in the previous section? To answer this question, we should see how the (IR-divergent) perturbative expansion in $1/\tau$ (equivalent to a perturbative expansion in $g_{st}$) is implemented in our present effective field theory setting.

Let us turn back to the action (\ref{DBI_D1}). Since the kinetic term for the $Y$-field is accompanied by the D1-brane tension $\tau$, the powers of $Y$ entering the various matrix elements contributing to the amplitude will be translated into powers of $1/\tau$. Let us concentrate for a moment on the case when the initial and final states of the D1-brane are the vacuum, since this is the amplitude we have studied in the string theory context. Then, the lowest order contribution (in $1/\tau$) to the amplitude (\ref{D1amplitude}) is
\beq
\left\langle 0,k_2\right|\frac{\tau}4\int dt dx Y^i(t,x)Y^j(t,x)\frac{\del^2}{\del y^i\del y^j}\left(\Phi(t,x,0)\right)^2\left|0,k_1\right\rangle .
\label{loop}
\eeq
This is a divergent expression proportional to
\beq
(k_1-k_2)^2\int \frac{d\kappa}{|\kappa|}.
\label{div}
\eeq
We immediately recognize the same dependence%
\footnote{The analog of $\left\langle V^{(1)}\cdots V^{(n)}\right\rangle _{D_2}$ of (\ref{annulus}) in our present computation is 
\be
\left\langle 0,k_2\right|\frac12\int dt dx \left(\Phi(t,x,0)\right)^2\left|0,k_1\right\rangle ,
\ee
which according to \eq{phisquared} is only non-zero when no Neumann momentum is transferred, and is independent of the transferred Dirichlet momentum.
} 
on the transferred Dirichlet momentum $(k_1-k_2)^2$ as we found for the annular divergence (\ref{annulus}). In fact, (\ref{loop}) is nothing but the contribution from the particular process we are considering to the string annular diagram. The annular diagram is divergent, and so is our present contribution. (Note that one needs to be careful in relating the cut-off parameter used
to regularize the divergent integral in (\ref{div}) to the world-sheet cut-off employed in string theory.)

Also note that when comparing the CFT divergence with the divergence in the $1/\tau$ expansion of the DBI result, the initial and final states in the DBI computation were chosen to be the vacuum state of the scalar fields $Y^i$. In this state, the expectation value of $(Y^i)^2$ diverges due to long wavelength contributions, so quantum fluctuations effectively delocalize the D1-brane completely in the Dirichlet directions. The quantum mechanical picture of D1-branes thus differs dramatically from the classical picture, where we had a well-localized D1-brane (which could develop a specific wave pattern under local recoil).%

What happens to the divergence \eq{loop} (in the context of effective field theory) when higher-order corrections are included? In fact, we have already derived the resummed expression (\ref{inprodcoh}). If all the powers of $1/\tau$ are kept in the expression of the amplitude (\ref{D1amplitude}), the result vanishes if the initial and final states of the D1-brane are the vacuum. This is in accord with momentum conservation.

Moreover, if we include more general final states of the D1-brane, we notice that the amplitude can only be non-vanishing for final states of the D1-brane that carry the amount of Dirichlet momentum dictated by momentum conservation. This is the local recoil phenomenon we have described in the introduction.
The discussions presented in this section can be straightforwardly
generalized to the case of higher-dimensional D-branes with
only one non-compact spatial direction.

\section{Relation to kinematics of low-dimensional field theories}

It has been repeatedly emphasized throughout the above presentation that the infrared divergences of the string perturbative expansion 
in the presence of a D-brane are
related to momentum conservation. This statement may seem paradoxical, since momentum is conserved for all possible D-branes, yet the divergences are present only for D$p$-branes with $p<2$.

One can furthermore examine, at the level of the DBI action, the higher-dimensional analog of the expression (\ref{inprodcoh}), derived in the previous section for the case of the D1-brane, and observe, for the case of higher-dimensional branes, that the transition amplitude is non-vanishing, for example, when both initial and final state of the D-brane are chosen to be vacuum states. Does it mean that momentum conservation is compromised?

The answer to the above question is most certainly no. However, the resolution of the paradox is instructive and somewhat subtle. One just needs to be conscious about what states of the D-branes appear in the CFT amplitudes. And this is, in turn, related to the kinematic properties of free massless scalar field theory in various dimensions.

Let us start with the intuitively straightforward case of the D0-brane.
In this case, the dynamics of the D0-brane is 
described by a free massless (0+1)-dimensional scalar field $X^i(t)$, which is simply the position of the D-brane. Momentum conservation is due to the translational symmetry of $X^i(t)$, and the scattering states of the D0-brane are momentum eigenstates. The scattering amplitudes contain a momentum conservation $\de$-function, and, as will be explained in \cite{CEN}, the worldsheet CFT attempts to expand this $\de$-function in a Taylor series, thereby producing the infrared divergences.

Imagine now we are to examine closed string scattering off a D$p$-brane with $p\ge 2$. Again, the deformations of its worldvolume are described by a free massless ($p+1$)-dimensional scalar field $X^i(\xi,t)$, and (Dirichlet) momentum conservation is due to the field translation symmetry $X^i\to X^i+a^i$. However, this symmetry is spontaneously broken. The ground state of a
Dp-brane with $p\ge 2$ is localized in the Dirichlet directions and does not have a definite value of the Dirichlet momentum. This explains why, even though momentum is conserved irrespectively of which D-brane one works with, for higher-dimensional D-branes, the amplitude for the D-brane to remain in its vacuum state (as the closed strings scatter off it) is non-vanishing, the general scattering amplitude does not contain a momentum conservation $\de$-function, and the infrared divergences are absent in the worldsheet CFT.

Why doesn't the case of the D1-brane fall into the same category as the higher-dimen\-sio\-nal branes? The deformations of a D1-brane worldsheet are once again described by a free massless (1+1)-dimensional scalar field $X^i(\xi,t)$, and (Dirichlet) momentum conservation is due to the field translation symmetry $X^i\to X^i+a^i$. However this symmetry is not spontaneously broken, since, according to a well-known result by Coleman \cite{coleman,marajaraman}, spontaneous breaking of continuous symmetries is impossible in two dimensions. The ground state of a D1-brane is not localized in the Dirichlet directions, and it does have a definite value of the Dirichlet momentum. This explains the discontinuous momentum dependence of the scattering amplitudes and the associated infrared divergences in the worldsheet CFT. 

\section*{Acknowledgments}
We would like to thank J. Ambj\o rn for collaboration in the early stages of the project. We also thank J.~de Boer, N.~Dorey, M.~Gaberdiel, M.~Green, V.~Hubeny, H.~B.~Nielsen, P.~Olesen, J.~Preskill, M.~Rangamani and D.~Tong for useful discussions. The work of B.C.\ was supported in part by Stichting FOM, by the Belgian Federal Science Policy Office through the Interuniversity Attraction Pole P5/27, by the European Commission FP6 RTN programme MRTN-CT-2004-005104 and by the ``FWO-Vlaanderen'' through project G.0428.06. S.N.\ was supported in part by the SRC Program of the KOSEF through the Center for Quantum Space-time (CQUeST) of Sogang University with grant number R11 - 2005 - 021. S.N.\ also thanks the Niels Bohr Institute where a part of the present work was done.


\appendix

\section{Local recoil of field-theoretic vortex lines}
In describing the phenomenon of local recoil, we emphasized its kinematic nature and hence its universality. In particular, local recoil should be manifest when particles scatter off a field-theoretic vortex line.

For the string theory example of local recoil that has been the main subject of our present paper, we have given a fairly complete formal account: starting with the infrared divergence in the loop diagram, we proceeded to show that the divergence is cancelled if an appropriate final state (satisfying momentum conservation) is chosen. We have further demonstrated, within the context of low-energy effective field theory, that, if one resums the infrared divergences to all orders in the string coupling, one reconstructs the (non-analytic) $\de$-function-like dependences on momenta typical of momentum-conserving processes.

For vortex lines in a field theory context \cite{vilenkin}, we shall not pursue the cancellation of divergences in a systematic fashion. It is nevertheless instructive to show that infrared divergences are still present in loop diagrams. These divergences indicate the relevance of the local recoil phenomenon for field-theoretic vortex lines, and prompt whatever further considerations of the divergence cancellation one might be willing to undertake.

Let us consider (for the sake of simplicity) vortex lines made of scalar fields. By Derrick's theorem, such string-like objects can only exist in 2+1 dimensions, and they are (2+1)-dimensional lifts of solitonic kink solutions of (1+1)-dimensional scalar field theories. One can consider, for instance, a (2+1)-dimensional real scalar field $\phi$ with action
\beq
S=\int d^3x\left\{\frac12\,\del^\mu\phi\,\del_\mu\phi-\frac{\lambda}4\left(\phi^2-\eta^2\right)^2\right\}.
\label{phi4}
\eeq
The equations of motion admit a static string-like solution
\beq
\phi=\phi^{(0)}(y)
\eeq
with $\phi^{(0)}(y)\to -\eta$ at $y\to-\infty$ and $\phi^{(0)}(y)\to \eta$ at $y\to+\infty$. This solution describes an infinite string with finite energy per unit length stretched in the $x$-direction.

If one performs quantization around such a classical solution (with $\phi=\phi^{(0)}+\varphi$), the action takes the form
\beq
S=\int d^3x\left\{\left(\del_t\varphi\right)^2-\left(\del_x\varphi\right)^2-\left(\del_y\varphi\right)^2+U(y)\varphi^2+{\cal L}_{int}(\varphi)\right\},
\label{deltaphi}
\eeq
where the exact expression for $U(y)$ can be determined from the solution $\phi^{(0)}$ (and its independence on $x$ and $t$ follows from the analogous independence of $\phi^{(0)}$). Furthermore, all the terms cubic and quartic in $\varphi$ have been symbolically assembled in ${\cal L}_{int}$.

Since ${\cal L}_{int}$ is proportional to the coupling constant $\lambda$, it can be treated perturbatively. Whenever one tries to compute perturbative corrections, say, to scattering amplitudes, they will contain the propagator, which is the inverse of the kernel of the quadratic Gaussian form in (\ref{deltaphi}), with the appropriate $i0$ prescription:
\beq
\ber{c}
\dsty\left(-\del_t^2+\del_x^2+\del_y^2+U(y)\right)G(t-t',x-x',y,y')=\de(t-t')\de(x-x')\de(y-y'),\vspace{3mm}\\
\dsty G(t-t',x-x',y,y')=\sum\limits_{\alpha}\frac{\Phi_\alpha(t,x,y)\Phi^*_\alpha(t',x',y')}{\alpha+i0},
\eer
\label{propgtr}
\eeq
where $\Phi_\alpha$ are eigenfunctions of the wave operator associated with (\ref{deltaphi}):
\beq
\dsty\left(-\del_t^2+\del_x^2+\del_y^2+U(y)\right)\Phi_\alpha(t,x,y)=\alpha\,\Phi_\alpha(t,x,y).
\label{Phialpha}
\eeq

We shall now show that the propagator is divergent for any values of the arguments (in other words, the inverse of the wave operator does not exist). Hence, the radiative corrections will be divergent, and one will need to modify the static vortex line background to take local recoil into account.

To demonstrate the divergence of the propagator, we first notice that, as a consequence of translational invariance of the original action (\ref{phi4}), $\Phi_0\equiv\del_y\phi^{(0)}$ satisfies the linearized equation of motion: 
\beq
\dsty\left(\del_y^2+U(y)\right)\Phi_0(y)= 0.
\label{Phi0}
\eeq
We then construct the following family of eigenfunctions (\ref{Phialpha}):
\beq
\Phi_{\omega,k}(t,x,y)=e^{i(\omega t-kx)}\Phi_0(y).
\eeq
Because of (\ref{Phi0}), they satisfy (\ref{Phialpha}) with 
\beq
\alpha_{\omega,k}=\omega^2-k^2.
\eeq
Hence, they will give the following contribution into $G(t-t',x-x',y,y')$ of (\ref{propgtr}):
\beq
G_{bend}(t-t',x-x',y,y')=\Phi_0(y)\Phi_0(y')\int d\omega\,dk\, \frac{e^{i\omega(t-t')-ik(x-x')}}{\omega^2-k^2+i0}.
\eeq
The integral over $d\omega\,dk$ is nothing but the propagator of a (1+1)-dimensional massless scalar field, which is known to be divergent (for any values of the space-time arguments) due to the singularity in the integrand at small $\omega$ and $k$. Note that the divergence of the (1+1)-dimensional massless scalar field propagator is essential to Coleman's proof of the absence of spontaneous symmetry breaking in 1+1 dimensions \cite{coleman,marajaraman}. This absence of spontaneous symmetry breaking is, in turn, intimately related to the local recoil phenomenon, as was explained in the main text.

Overall, we have seen that, for quantum field theories in the background of a vortex solution, one discovers a set of modes which correspond to bending of the string. These modes possess the spectrum of (1+1)-dimensional massless scalar fields, and they give a divergent contribution to the propagators of the (space-time) fields of the fundamental Lagrangian. Because the propagators are divergent, so will be the radiative corrections. One therefore concludes that, to perform computations beyond leading order in the coupling, one will need to modify the background of a static vortex line, that is, to take into account the phenomenon of local recoil.

\end{document}